# Fourier Optics approach to imaging with sub-wavelength resolution through metal-dielectric multilayers

R. KOTYŃSKI[*]

University of Warsaw, Faculty of Physics
Pasteura 7, 02-093 Warsaw, Poland

Metal-dielectric layered stacks for imaging with sub-wavelength resolution are regarded as linear isoplanatic systems – a concept popular in Fourier Optics and in scalar diffraction theory. In this context, a layered flat lens is a one-dimensional spatial filter characterised by the point spread function. However, depending on the model of the source, the definition of the point spread function for multilayers with sub-wavelength resolution may be formulated in several ways. Here, a distinction is made between a soft source and hard electric or magnetic sources. Each of these definitions leads to a different meaning of perfect imaging. It is shown that some simple interpretations of the PSF, such as the relation of its width to the resolution of the imaging system are ambiguous for the multilayers with sub-wavelenth resolution. These differences must be observed in point spread function engineering of layered systems with sub-wavelength sized PSF.

# 1. Introduction

Until the introduction of the concept of a perfect flat lens with either a single layer [1-4] or with multiple layers [5], it was rather uncommon to regard multilayers as (spatial) imaging systems, or linear spatial filters. Instead, the usual characteristics of a multilayer includes the dependence of transmission and reflection coefficient on the angle of incidence and polarisation, as well as the photonic band structure in case of periodic stacks. However, in order to describe the resolution of an imaging system consisting of a multilayer in a systematic way, it is convenient to refer to the theory of linear shift invariant systems (LSI, also termed as linear isoplanatic systems [6,7]). In this paper, metal-dielectric multilayers (MDM) are regarded as LSI systems and a layered superlens is a one-dimensional spatial filter characterised with the point spread function (PSF). This approach may facilitate the application of plasmonic elements to optical signal processing which is currently bringing increasing research interest [8].

Since the seminal paper by Pendry [1] subwavelength imaging at visible wavelengths brought a large interest [9-29] and in particular has been demonstrated in much thicker low-loss layered silver-dielectric periodic structures [9-20]. A variety of physical models may be applied to explain the mechanism of transmission: 1. the effective medium anisotropic approximation of the sub-wavelength multilayer [9,27] combined with the Fabry-Perot resonant condition tuned to be independent of the angle of incidence [13-15]; 2. multiple negative refraction resulting in diffraction-free propagation [10]; 3. resonant tunnelling through the bandgap material formed by the periodic metal-dielectric multilayer [9].
The enhancement of evanescent fields needed for sub-wavelength imaging may be also explained in various ways: 1. as the result of collective coupling between plasmon modes at subsequent metallic layers [12,19] - if we look to the internal field distributions; 2. as self-collimation [29] - if we look to the band diagrams of the multilayer; 3. as the result of large effective permittivity $\varepsilon_\perp^{EMT} \gg 1$ - when the homogenized anisotropic model of the system is valid.

---

[*] E-mail: rafal.kotynski@fuw.edu.pl

In a recent letter [17], it was demonstrated that the properties of PSF of a layered superlens are different than those of typical PSFs that characterise classical imaging systems. For instance the image of a narrow sub-wavelength Gaussian incident field may be surprisingly dissimilar to the PSF, and the width of PSF is not a straightforward measure of resolution. FWHM or standard deviation of PSF give ambiguous information about the actual resolution, and imaging of objects smaller than the FWHM of PSF is sometimes possible. A multiscale analysis of imaging gives good insight into the peculiar scale-dependent properties of sub-wavelength imaging and provides the means to distinguish between diffraction-free propagation for various ranges of object sizes.

In the present paper a more thorough background and further discussion on the results reported in [17] are presented. For this reason, the same multilayer consisting of silver and Strontium Titanate will serve as an example used in the simulations presented in this paper. The main focus of the present paper is put on the description of the imaging system with use of the concepts borrowed from Fourier Optics. Furthermore, the distinction is made between a soft source and a hard electric and magnetic sources. Each of these definitions leads to a different meaning of perfect imaging.

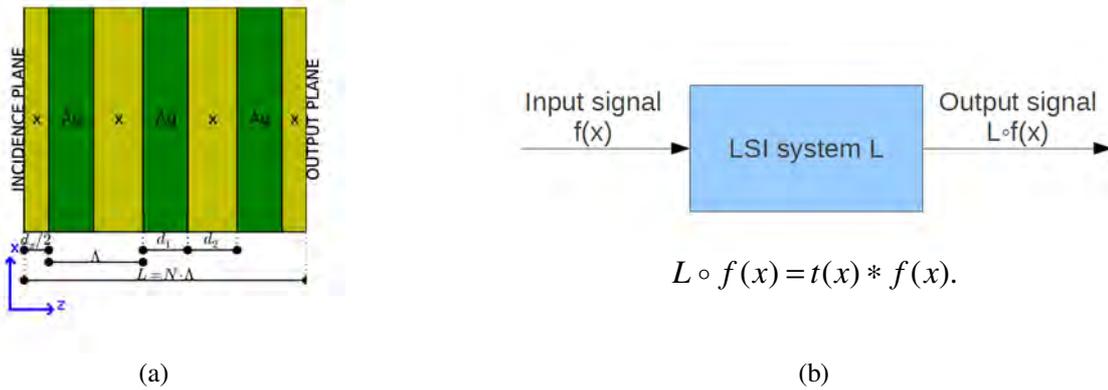

(a) (b)
**Fig. 1:** a) Schematic of a periodic multilayer consisting of silver and a dielectric x; b) Schematic of a linear shift invariant (LSI) system.

## 2. LSI imaging systems consisting of metal-dielectric multilayers

Figure 1a. shows a schematic of a periodic multilayer with silver and dielectric layers. We consider coherent imaging from its left-side boundary to the right side. The multilayer is suspended in air, or more generally in a dielectric material, which is the same at its both sides.

Let us recall the basic concepts and terminology related to linear shift invariant systems [6,7]. A scheme of an LSI system is presented in Fig. 1b. The system transforms the input signal into an output signal. Mathematically, a system is represented with an operator, while the signal is represented with a function. The system is said to be linear when the corresponding operator $L$ is linear and satisfies the following condition for any two input signals $f_1$ and $f_2$ and for any two scalar factors $\alpha_1$ and $\alpha_2$,

$$L \circ (\alpha_1 \cdot f_1 + \alpha_2 \cdot f_2) = \alpha_1 \cdot L \circ f_1 + \alpha_2 \cdot L \circ f_2. \qquad (1)$$

In optics, the input and output signals are usually defined as the field at certain locations of the optical set-up. It is common to consider a signal that is either temporal or spatial, scalar or vectorial, real or complex valued, and either one- or multi-dimensional. In our present analysis we consider monochromatic and spatially coherent imaging, hence the signal is defined with the spatial distribution of a complex amplitude representing a scalar component of electric or magnetic field, while the linearity of the system reflects the simple superposition principle and depends on the use of optically linear materials.

A linear spatial system is said to be shift-invariant (or isoplanatic) if $L$ commutes with the operator of translation, or in other words if the response of the system to a shifted signal, is shifted by the same distance but is otherwise unaffected,

$$L \circ (f(x) * \delta(x - x_0)) = (L \circ f(x)) * \delta(x - x_0), \qquad (2)$$

where $*$ stands for convolution.

Linear shift invariant systems (LSI) are precisely characterised by their respective point spread function (PSF, also referred to as the impulse response) $t(x) = L \circ \delta(x)$ since the response of the system to an arbitrary signal is equal to the superposition integral of the following simple form

$$L \circ f(x) = t(x) * f(x). \qquad (3)$$

Equivalently, the convolution theorem allows to rewrite the same formula in the Fourier domain

$$FT\{L \circ f(x)\} = FT\{t(x)\} \times FT\{f(x)\} = \hat{t}(k_x) \times \hat{f}(k_x), \qquad (4)$$

where $FT\{.\}$ as well as the dash represent the Fourier transform, and $\hat{t}(k_x) = FT\{t(x)\}$ is the transfer function of the LSI. The transfer function is therefore the ratio of the spatial spectra of the output to input signals, and the operation of LSI may be understood as spatial filtering.

Let us now consider a planar imaging system shown in Fig. 1a. and consisting of a multilayer suspended in air, with infinite and parallel layers composed of optically linear materials with isotropic but complex permittivities. For coherent, planar imaging, either for TM or TE polarisation, this systems is a scalar LSI. Linearity is the direct consequence of the superposition principle for coherent fields in linear optical materials. Shift-invariance is a natural property of a system which has no unique optical axis due to the assumed infinite size of the layers. Finally, the system is scalar, because all the components of the electric and magnetic fields may be calculated from the single component of the magnetic field $H_y(x, z)$ in case of the polarisation TM, and from the single component of the electric field $E_y(x, z)$ in case of the polarisation TE. Indeed, we have,

$$\begin{pmatrix} E_x(x,z) \\ E_z(x,z) \end{pmatrix} = \frac{i \cdot \eta_0}{k_0 \cdot \varepsilon(x,z)} \begin{pmatrix} \partial/\partial z \\ -\partial/\partial x \end{pmatrix} H_y(x,z), \quad \begin{pmatrix} H_x(x,z) \\ H_z(x,z) \end{pmatrix} = \frac{i}{k_0 \cdot \eta_0 \cdot \mu(x,z)} \begin{pmatrix} -\partial/\partial z \\ \partial/\partial x \end{pmatrix} E_y(x,z), \qquad (5)$$

where $k_0 = 2\pi/\lambda$ and $\eta_0 = \sqrt{\mu_0/\varepsilon_0}$ are the free-space wavenumber and impedance, respectively, and $\mu \equiv 1$ for non-magnetic materials. For the matter of simplicity, further we refer only to the TM polarisation. Notably, surface plasmon polaritons exist only for the TM polarisation, and this polarisation is more important for applications in superresolution. The one-dimensional spatial spectrum of the magnetic field, at every position $z$ is equal to,

$$\hat{H}_y(k_x, z) = \int_{-\infty}^{\infty} H_y(x, z) \cdot \exp(ik_x x) \cdot dx \qquad (6)$$

This spatial spectrum has a similar significance within a multilayer as it has in free space, because $H_y$ is continuous and $k_x$ is conserved at layer boundaries. Let us now further exploit the analogy with diffraction. Propagation of the spatial spectrum in free-space, is a convenient way to describe and model diffraction. The respective transfer function at a distance $L$ in air is equal to,

$$\hat{t}(k_x) = \frac{\hat{H}_y(k_x, z = L)}{\hat{H}_y(k_x, z = 0)} = \exp(ik_z L) = \exp\left(iL\sqrt{k_0^2 - k_x^2}\right) \tag{7}$$

In wave optics usually a scalar field approximation is assumed, and Eq. (7) is written in two dimensions $(x, y)$ for some scalar field $U$ with neglection of polarisation effects.

For completeness it should be mentioned that the Fresnel diffraction approximation results from the second order Taylor expansion of the phase of the transfer function (7) for propagating waves ($k_0^2 > k_x^2$) [6,7],

$$L\sqrt{k_0^2 - k_x^2} \approx k_0 L(1 - \theta^2/2), \tag{8}$$

where $\theta^2 = \lambda^2 \nu_x^2$ approximates the angle with the optical axis, and $\nu_x = k_x/2\pi$ is the spatial frequency. The near-field approximation was successfully applied to some metal-dielectric layered systems as well [9].

Now, let us return to the imaging LSI system based on a multilayer. Due to reflections from the layer boundaries, within the multilayer and at its left-hand side, the spatial spectrum contains contributions from planewaves propagating both in the forward and backward directions,

$$\hat{H}_y(k_x, z) = \alpha^+_{z=z_0} \cdot \exp(+i(z-z_0) \cdot k_z) + \alpha^-_{z=z_0} \cdot \exp(-i(z-z_0) \cdot k_z), \tag{9}$$

where $\alpha_\pm$ are the planewave amplitudes within a layer, and $k_z = \sqrt{\varepsilon \cdot k_0^2 - k_x^2}$. The spatial spectrum of the electric field $\hat{E}_x(k_x, z)$ is now obtained using Eq. (5),

$$\hat{E}_x(k_x, z) = \beta^+_{z=z_0} \cdot \exp(+i(z-z_0) \cdot k_z) + \beta^-_{z=z_0} \cdot \exp(-i(z-z_0) \cdot k_z) \text{ where } \beta^\pm = \mp \frac{k_z \cdot \eta_0 \cdot \alpha^\pm}{k_0 \cdot \varepsilon}, \tag{10}$$

The contributions from planewaves propagating in the forward and backward directions in equations (9) and (10) have a different sign, $\beta^\pm \propto \mp \alpha^\pm$. As it will be shown, this results in a different form of the transfer function and point spread function for a hard electric source, and hard magnetic source.

The transfer matrix method [34] is used to determine the amplitudes $\alpha^\pm$ or $\beta^\pm$ within every layer of the stack and to calculate the transfer function as,

$$\hat{t}(k_x) = \frac{\hat{H}_y(k_x, z=L)}{\hat{H}_y^{in}(k_x, z=0)} = \frac{\hat{E}_x(k_x, z=L)}{\hat{E}_x^{in}(k_x, z=0)} = \frac{\alpha^+_{z=L}(k_x)}{\alpha^+_{z=0}(k_x)} = \frac{\beta^+_{z=L}(k_x)}{\beta^+_{z=0}(k_x)}, \tag{11}$$

where $\hat{E}_y^{in}(k_x, z=0)$, $\hat{E}_x^{in}(k_x, z=0)$ are the incident magnetic and electric fields.

Equation (11) is the most natural definition of the transfer function. Further this definition will be termed as the transfer function for a soft source model, since its definition completely neglects reflections. In electromagnetic modelling, and in particular in FDTD, a soft source is a popular model of the source which is defined in such a way that it introduces the incident electromagnetic wave at some point or area of the computational domain, but does not interact with nor block the reflected waves. Equation (11) indicates that $\hat{t}(k_x)$ has the same form for electric and magnetic field. Therefore, the corresponding point spread functions resulting from a point magnetic or electric (soft) sources are also the same.

In this paper two other possible definitions of the transfer function are also proposed,

$$\hat{t}'(k_x) = \frac{\hat{H}_y(k_x, z=L)}{\hat{H}_y(k_x, z=0)} = \frac{\alpha^+_{z=L}(k_x)}{\alpha^+_{z=0}(k_x) + \alpha^-_{z=0}(k_x)} = \frac{\beta^+_{z=L}(k_x)}{\beta^+_{z=0}(k_x) - \beta^-_{z=0}(k_x)}, \quad (12)$$

and

$$\hat{t}''(k_x) = \frac{\hat{E}_x(k_x, z=L)}{\hat{E}_x(k_x, z=0)} = \frac{\beta^+_{z=L}(k_x)}{\beta^+_{z=0}(k_x) + \beta^-_{z=0}(k_x)} = \frac{\alpha^+_{z=L}(k_x)}{\alpha^+_{z=0}(k_x) - \alpha^-_{z=0}(k_x)}. \quad (13)$$

These definitions refer to the hard magnetic source and hard electric source, respectively, since they relate the outgoing field at the output plane to the total field at the input plane. A hard (fixed) source is a popular model of a source used e.g. in FDTD, in which the total field at a certain point or area of the simulation domain is assumed to be known *a priori*. The hard source, as it is defined here, is similar but not equivalent to a hard source used in FDTD. For instance, in FDTD a hard source with a finite spatial size may be responsible for the scattering of the reflected wave. Here, speaking of a hard source we assume that the total field in the entire incidence plane is known, therefore its spatial spectrum is known *a priori* as well, which is usually not the case in FDTD. Nevertheless, the present definition is compatible with the hard sources used in FDTD in the sense that it represents a source separated from the computational domain with a plane which is perfectly reflecting from one side and perfectly transmitting from the other. This property is proven in the Appendix. Multiple reflections between the multilayer and a hard source are therefore properly accounted for.

Finally, it should be emphasized that a realistic physical near-field source is likely to interact with the reflected wave in a more complex way than the hard and soft sources considered here. The source models are chosen due to their simple form adequate for the use in numerical modelling using TMM or FDTD, when it is not possible to include the real source in the simulations. The three models analysed in this paper represent the sources which are non-reflecting, perfectly reflecting for the electric field, and perfectly reflecting for the magnetic field, respectively, while any real source is likely to be partly reflecting and therefore would represent an intermediate situation. Nevertheless, it is possible to argue rather qualitatively that a plane-wave diffracted on a phase mask resembles a soft source, as the reflected wave may freely propagate in the backward direction. Conversely, a plane-wave diffracted on a mask made of a perfect metal with narrow slits resembles a hard electric source, as a large amount of reflected energy is once again reflected towards the multilayer by the boundary conditions.

While it is convenient to characterise imaging through a multilayer using the framework of LSI, there are several limitations of this model which must be observed. They are due to reflections, the presence of the source (mask, fiber tip etc) in the near-field, and the need to include evanescent waves in the spatial spectrum. Let us summarise this section with a discussion on the properties of the transfer function specific to multilayers.

1. The scalar LSI model applies to the planar imaging with the TE and TM polarisations. However, full 3D imaging of two-dimensional images involves the coupling between TE and TM polarisations and requires a fully vectorial approach. In such a situation, it is necessary to generalise the PSF to a take a matrix form [31].
2. The transfer function is defined as the ratio of the output to input spatial spectra. For the TM polarisation we have, $\hat{t}(k_x) = \hat{H}_y^{out}(k_x)/\hat{H}_y^{in}(k_x)$. However, due to reflections, there is an ambiguity in the definition of the "input" field – one may choose between the incident field (part of which is transferred or absorbed and part of which is reflected), and the total field at the incidence plane (resulting from the interference between the incident and reflected light). Definition of the transfer function in Eq. (11) corresponds to the first possibility, while the definitions in Eqs. (12) and (13) to the latter. For the propagating waves or for a far-field source it is the most natural to define the "input" field only with the plane-waves propagating towards the multilayer. The natural extension of this definition to evanescent fields is to assume that a field decaying with distance from the source contributes to the incident field, while the field decaying in the opposite direction is the reflected field. This reasoning leads to definition (11). On the other hand, for evanescent waves, it is a matter of convention to distinguish between the incident and reflected wave, and this is an argument to define the "input" field as the total field, resulting in definitions (12) or (13).

3. The significance of PSF is limited to a selected scalar field component, e.g. $H_y$. Other field components may be calculated using (5) and usually have a dissimilar shape than $H_y$.
4. The width of the PSF is not always a simple measure of the resolution of the system. This point will be further discussed in the next sections. The interpretation of a PSF of a multilayer is therefore less straightforward than in most classical LSI imaging systems.

## 3. Multiscale analysis of resolution

The popular Rayleigh criterion of resolution assumes that the images of two incoherent point sources may be resolved, if the centre of one image coincides with the first minimum of the other one. This minimum separation depends on the wavelength and the numerical aperture *NA* and is equal to $\delta_R \approx 0.61\lambda/NA$. However, the same criterion of resolution applied to coherent imaging becomes dependent on the phase shift between the two images [7]. Depending on the phase shift, the resulting resolution is either better or worse as compared to the case of incoherent imaging. Up to date, there is probably no standard and generally accepted resolution criterion precisely defined for coherent imaging with sub-wavelength resolution.

PSF of an LSI imaging system can be often straightforwardly interpreted, and provides complete information about the resolution, loss or enhancement of contrast, as well as the characteristics of image distortions. This information may be usually simply and straightforwardly extracted from the shape of PSF. For instance, the resolution may be usually linked to the width of PSF.

The support of a function $f: X \to Y$ is the subset of its domain $X$ where the function takes non-zero values $supp(f) = \{x \in X : f(x) \neq 0\}$. When the input signal and PSF are non-negative real-valued functions ($PSF(x), H_y(x) \in R^+$) and each of their supports forms an (open or closed) interval $\overline{supp(f)} = [x_1, x_2] \subset X$, the support of their convolution is also an interval. Moreover, the lengths of supports simply add together, contributing to the broadening of the filtered signal,

$$\overline{\overline{supp(H_y(x) * PSF(x))}} = \overline{\overline{supp(H_y(x))}} + \overline{\overline{supp(PSF(x))}}, \qquad (14)$$

Here, $\overline{supp(f)}$ denotes the closure of the support, and $\overline{\overline{supp(f)}}$ denotes its length. Defining the resolution $\delta$ as a measure of broadening of the filtered signal we may write, $\delta = \overline{\overline{supp(PSF(x))}}$. On the other hand, for simple Gaussian PSF and input, the output has the width (variance) equal to the sum of variances of PSF and input,

$$\exp(-x^2/\sigma_1^2) * \exp(-x^2/\sigma_2^2) \propto \exp(-x^2/(\sigma_1^2 + \sigma_2^2)). \qquad (15)$$

Therefore once again the width of PSF has a clear link to the resolution of the imaging system. However, formulas (14) and (15) are not a good reference for diffractive systems with complex-valued PSF. Nevertheless, also then, the width of PSF provides some qualitative information on the resolution of the system. For instance, when the input signal and PSF are complex-valued and their supports are bounded, Eq. (14) turns from an equality to an inequality ($\leq$) and provides an upper bound for the resolution, $\delta \leq \overline{supp(PSF(x))}$.

For the purpose of the present analysis of an LSI system, it is convenient to take the width of the PSF as a measure of resolution. This width may be expressed using full width-at-half-maximum $FWHM\left(|PSF|^2\right)$ or the standard deviation of $\sigma\left(|PSF|^2\right)$ which characterise the size of an intensity spot resulting from the image of a point object. Notably $FWHM\left(|PSF|^2\right)$ provides the information of the size of the central spot, while $\sigma\left(|PSF|^2\right)$ is very sensitive to the side-lobes and to the asymptotic behaviour of $|PSF(x)|^2$ further from the centre. These two criteria will be used in the next section, where their dependence on the width of a spatial input Gaussian signal is analysed. This is a way to conduct multiscale analysis of the resolution of an LSI multilayer imaging. Notably, the resolution depends on the width of the input signal, as well as the type of source, even though the PSF is independent of the size of the incident beam.

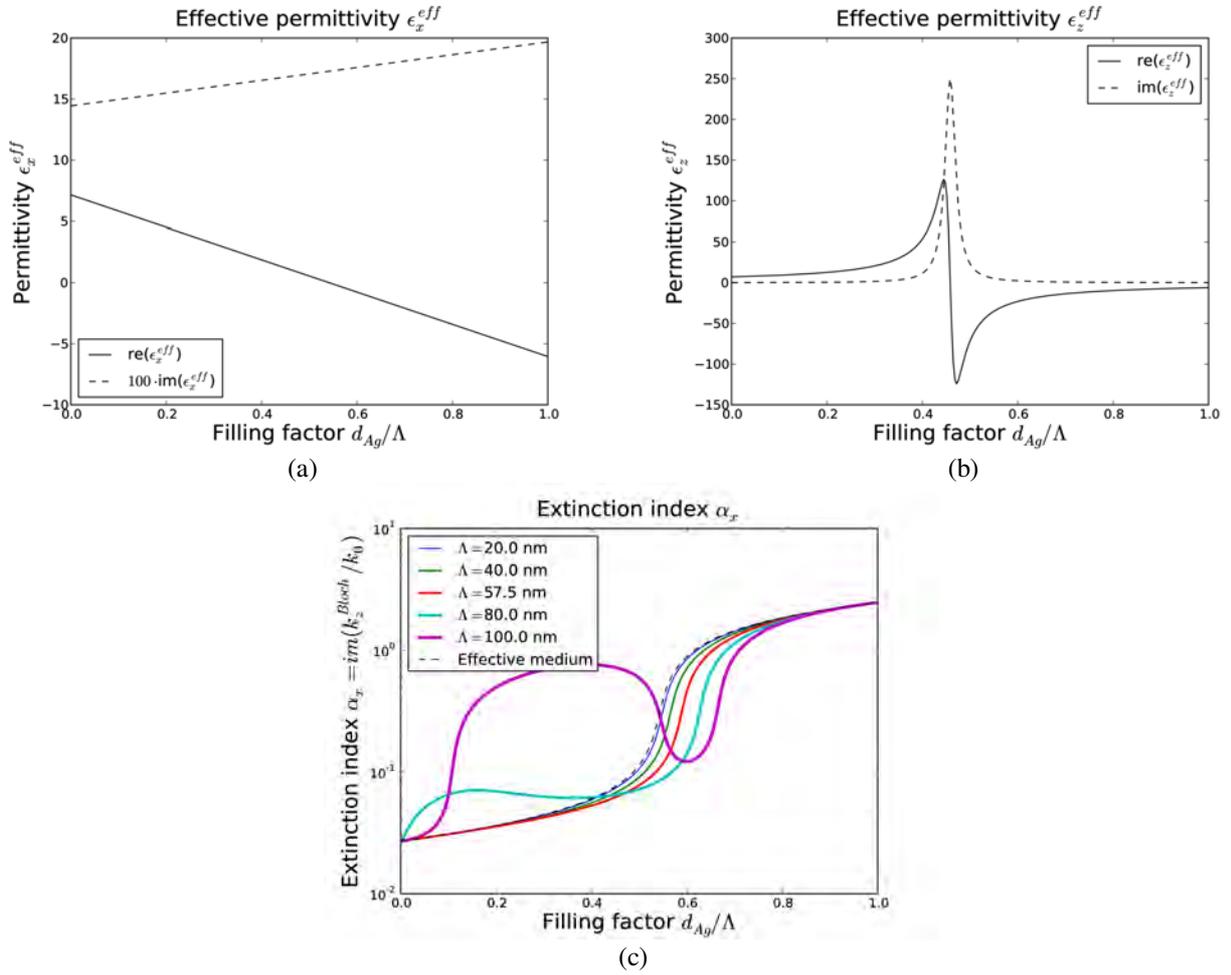

Fig 2: a,b) Effective permittivity $\varepsilon_x^{eff}$ (a) and $\varepsilon_z^{eff}$ (b) of the Ag-SrTiO$_3$ superlens at the wavelength of $\lambda = 430 nm$ as a function of the filling factor. c) extinction index $\alpha_x$ calculated using the effective medium model, as well as from the imaginary part of the Bloch wavenumber in an infinite periodic stack for various values of the lattice pitch $\Lambda = 20 \div 100 nm$.

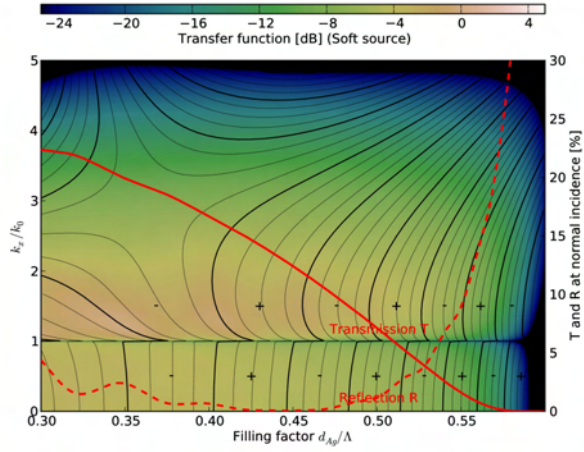

(a)

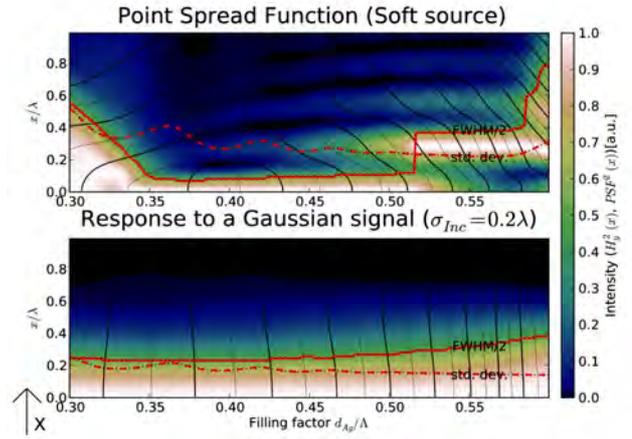

(b)

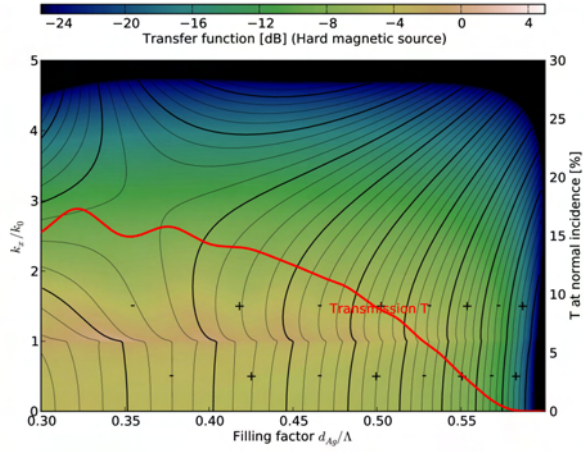

(c)

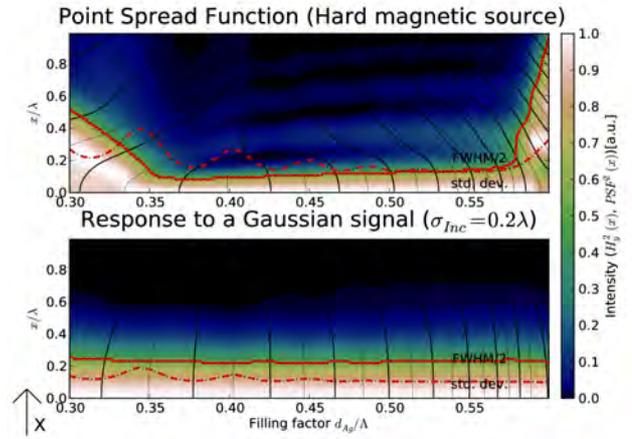

(d)

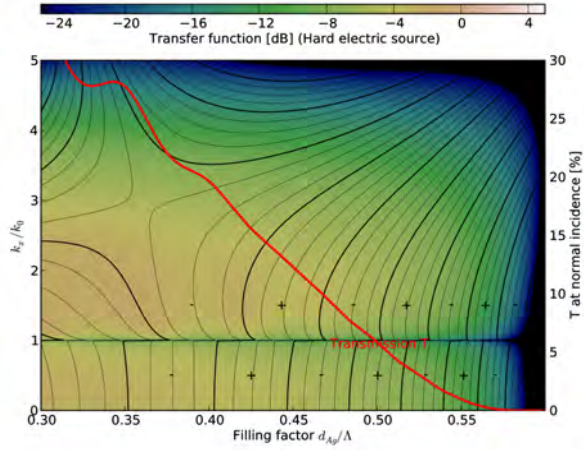

(e)

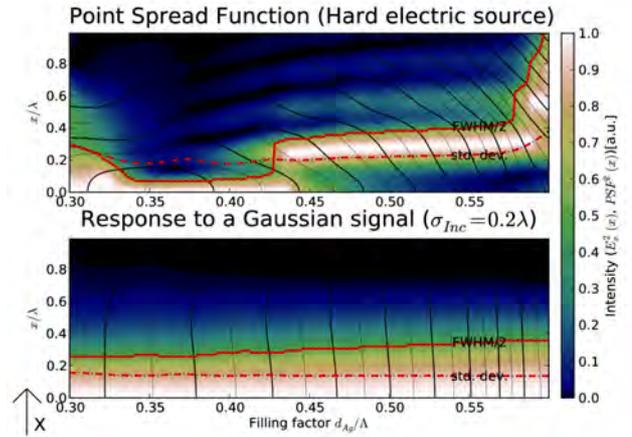

(f)

Fig. 3: Imaging properties of the multilayer for a soft source (a,b) hard magnetic source (c,d) and hard electric source (e,f) as a function of the filling factor of silver; a,c,e) The transfer functions $\hat{t}(k_x)$, $\hat{t}'(k_x)$, $\hat{t}''(k_x)$, are drawn in vertical cross-sections of the respective plots. Amplitude of the transfer function is shown in dB, and the phase is presented with the isolines separated by $\pi/4$ with the marked sign of the real part of amplitude. Intensity transmission and reflection coefficients at normal incidence are overdrawn on the transfer function. b,d,f) Point spread function of the multilayers drawn in vertical cross-sections as a function of the filling factor of silver together with the response to a narrow subwavelength Gaussian wavefront. Phase isolines are separated by $\pi/2$.

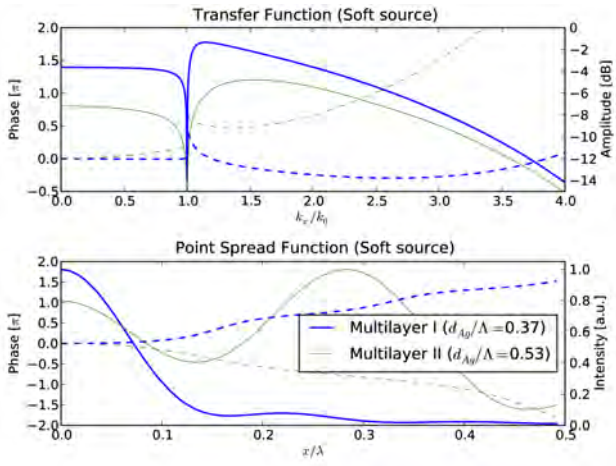
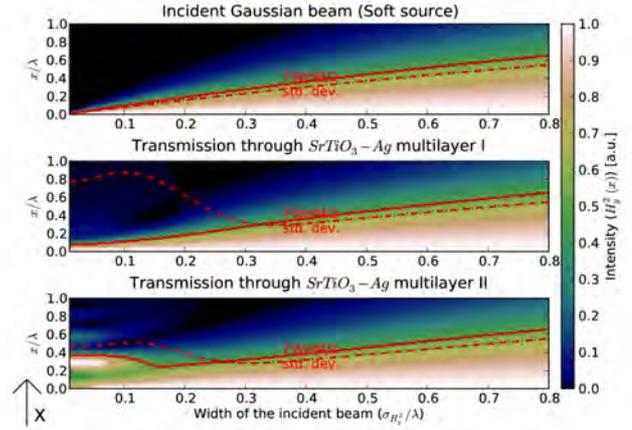

(a)                                                    (b)

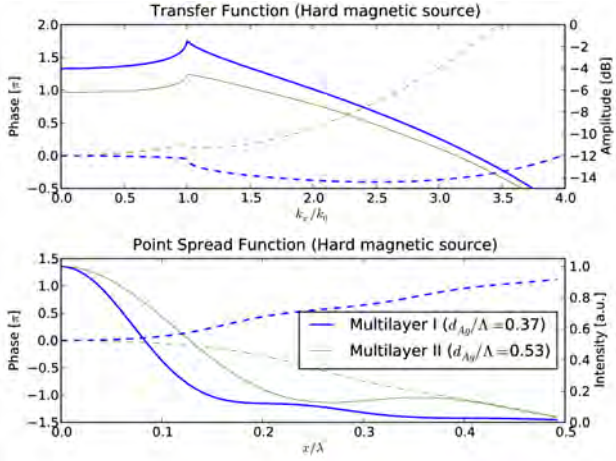
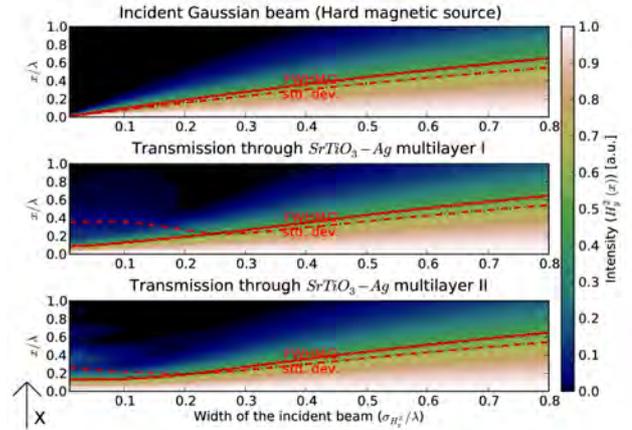

(c)                                                    (d)

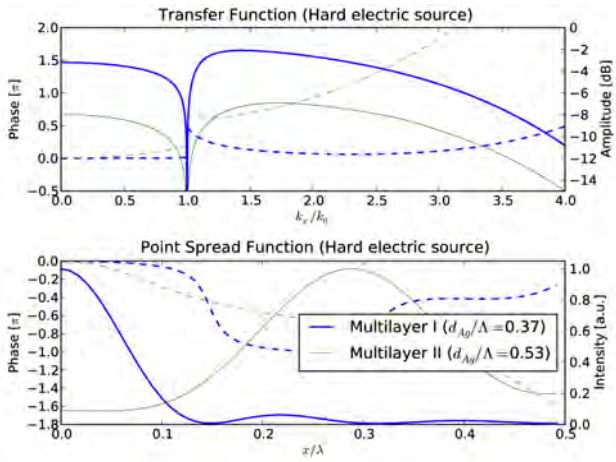
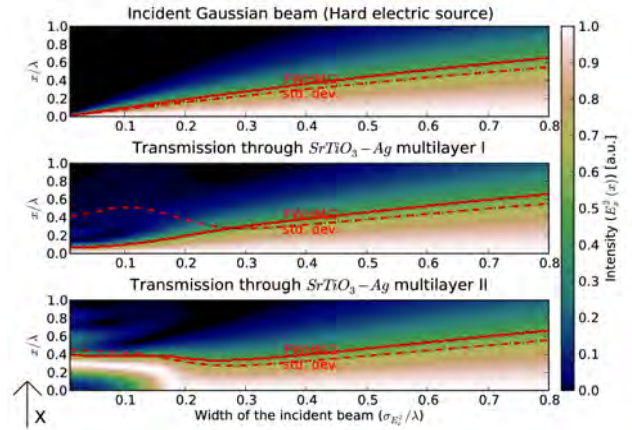

(e)                                                    (f)

Fig. 4: Comparison of the imaging properties of the multilayer I with $d_{Ag}/\Lambda = 0.37$ and multilayer II with $d_{Ag}/\Lambda = 0.53$ for a soft source (a,b) hard magnetic source (c,d) and hard electric source (e,f); a,c,e) The transfer functions and PSF of the multilayers; b,d,f) Response to a narrow subwavelength Gaussian incident wavefront of two multilayers. The responses as well as the incident wavefront are drawn in vertical crossections, vs. the width of the incident wavefront.

## 1. Numerical results

From now on, we focus on a SrTiO$_3$ – Ag multilayer with $N = 20$ periods, and with the total thickness $L = 1150nm$ [17]. The filling factor of silver $d_{Ag}/\Lambda$ is taken as a free parameter. The elementary cell consists of an Ag layer symmetrically coated with SrTiO$_3$ – this shape of the elementary cell is in agreement with the scheme in Fig. 1a and results in thinner external dielectric layers of the stack, as compared to the internal dielectric layers. Strontium Titanate is an isotropic material with a high refractive index $n = 2.674 + 0.027i$ at the wavelength $\lambda = 430nm$ [32]. The refractive index of silver at the same wavelength is equal to $n_{Ag} = 0.04 + 2.46i$ [33].

Some insight into the dependence of the imaging properties of the stack on the filling factor may be obtained using the effective medium model. However, the assumed lattice pitch equal to $\Lambda = L/N = 57.5nm$ is too large for the effective medium model to provide a satisfying quantitative description of the multilayer's operation. In particular, the effective medium model significantly overestimates the losses of the structure. After homogenisation, the multilayer may be modelled as a uniform slab made of an uniaxially anisotropic effective material [9]. The effective permittivity of the slab vs. filling factor is plotted in Figs. 2ab. For the filling factor of $d_{Ag}/\Lambda \approx 0.45$, the multilayer becomes approximately impedance-matched with air $\varepsilon_x^{eff} \approx 1$ and the large magnitude of $|\varepsilon_z^{eff}| \gg 1$ makes its dispersion equation for the TM polarised light,

$$\frac{k_z^2}{\varepsilon_x^{eff}} + \frac{k_x^2}{\varepsilon_z^{eff}} = k_0^2 \qquad (16)$$

almost independent of the spatial frequency $k_x$ resulting in diffraction-free propagation [9,14,18]. Therefore, the reflections are mitigated thanks to the condition of impedance matching, while the second condition $|\varepsilon_z^{eff}| \gg 1$ enables superresolving imaging (including the evanescent harmonics of the partial spectrum $k_x^2 > 1$) at distances limited by the losses. It should be noted that for $L = 1150nm \approx n_{SrTiO_3} \cdot \lambda$ the Fabry-Perot resonances are shallow and the actual thickness of the structure $L$ is of secondary importance for the transmission properties of the stack. The effective extinction coefficient of the structure is presented in Fig. 2c. It is first calculated using the effective medium model, and then for an infinite periodic stack with a finite value of the lattice pitch $\Lambda = 20 \div 100nm$. The latter is calculated from the eigenvalues of the transfer matrix for a single period of the structure. This comparison indicates that due to the steep slope of slope $\alpha_x(d_{Ag}/\Lambda)$ in the vicinity of $d_{Ag}/\Lambda \approx 0.45$, the effective medium model overestimates the losses already for such small values of the lattice pitch as $\Lambda \approx \lambda/10$. All further calculations are based on the transfer matrix method, which does not depend on the effective medium approximation. In [17], the superresolving properties of the multilayer were also verified with FDTD simulations.

Figure 3 presents a short summary of the imaging properties of the multilayer as a function of the filling factor, analysed for the three variants of the incident conditions - namely for a soft source (top), hard magnetic source (middle) and hard electric source (bottom). The subplots show the transfer function (Fig. 3a,c,e), the point spread function (top subplots in Figs. 3b,d,f), and the response to a narrow ($\sigma_{Inc} = 0.2\lambda$) Gaussian input signals (bottom subplots in Figs. 3b,d,f). All these functions have an even symmetry, therefore the domain in the plots is limited to $k_x \geq 0$ or $x \geq 0$, respectively. The condition of impedance matching may be seen as the main reason of the efficient removal of reflections for $d_{Ag}/\Lambda \approx 0.45$ (see the dependence of $R(d_{Ag}/\Lambda)$) presented in Fig. 3a). However, the reflections do exist for higher spatial frequencies, and therefore the shape of PSF at $d_{Ag}/\Lambda \approx 0.45$ is not the same for the three models of the source.

The evanescent part of the transfer function ($k_x > 1$, Figs. 3a,c,e) has a large magnitude, which is the necessary condition for sub-wavelength imaging. The shape of transfer function is generally regular with the exception of the phase discontinuity in the vicinity of $k_x/k_0 = 1$, as well as the strong phase modulation below $d_{Ag}/\Lambda < 0.35$ which suppresses the super-resolving properties of the PSF in that range. The phase step at $k_x/k_0 = 1$ in the transfer function influences the shape of the corresponding PSF which, with the increase of filling factor, evolves from a narrow sub-wavelength maximum to a shape dominated by the side-lobe. The response to a narrow sub-wavelength Gaussian signal is entirely different from the PSF (bottom subplots in Figs. 3b,d,f). PSF does not resemble a Gaussian function and its width measured with FWHM is different from the doubled standard deviation.

The off-axis background of PSF results in the high value of std. dev., and probably FWHM is a more meaningful measure of resolution of the system. Moreover, the broadening of the optical signal can not be expressed with formulas (14) or (15). The exception is the range of filling fraction in between 0.35 and 0.45, where the PSF resembles a Gaussian function and the broadening follows a simple intuitive behaviour.

More in general, the width of response may even show an anomalous (decreasing) dependence on the size of the sub-wavelength Gaussian incident signal. This effect is even more evident from the further simulations presented in Fig. 4. It is striking how dissimilar are the PSF and the response to a narrow Gaussian signal around $d_{Ag}/\Lambda \approx 0.5$ (See Figs. 3b,d,f). The explanation is nevertheless not difficult, as the bandwidth of the TF in use depends (inversely) on the width of the incident Gaussian function. The opposite phase of transfer function for propagating and evanescent waves is the source of the side-lobe of PSF. Partial removal of the central maximum in PSF (equal to the mean value of TF) occurs only when the contribution from evanescent and propagating harmonics to the mean value compensate each other. Broader Gaussian incident fields limit the bandwidth in use, and suppress this sensitive condition.

Figure 4 presents the transfer function, and point spread function of two selected multilayers with the filling fractions equal to $d_{Ag}/\Lambda \approx 0.37$ for multilayer I, and equal to $d_{Ag}/\Lambda \approx 0.53$ for multilayer II (see Figs. 4a,c,e), as well as their response to a sub-wavelength Gaussian field distribution with $FWHM < 1.6\lambda$ (see Figs. 4b,d,f). For a soft source, these two multilayers represent the situation of a regular nearly Gaussian PSF and a side-lobe dominated PSF, respectively. Both multilayers allow for imaging of subwavelength details, however their responses scale differently with the size of sub-wavelength object. Moreover, due to larger reflections, the imaging properties of multilayer II change considerably for different incident conditions. In fact, for a hard magnetic source, the PSF of this multilayer is no longer side-lobe-dominated. Different behaviour of multilayers I and II may be understood as resulting from the different value of the phase shift between the propagating and evanescent part of the transfer functions for the two multilayers (Figs. 4a,c,d), although this reasoning is only qualitative. We have recently analysed an analogous situation [16], however resulting from the different modulation depth of TF.

Let us still look to the multiscale anlysis of resolution presented in Figs. 4b,d,f. From these figures it is possible to determine the range of object dimensions which are imaged without distortions through the superlens. In this situation imaging resembles a diffraction-free projection of the incident image (see the FDTD simulations in [17]). While multilayer I allows for approximately diffraction-free propagation, independent of the size and type of the source, multilayer II behaves in the same way for broader sources only and shows strong diffraction when the shape of the source approaches a $\delta$-function. Moreover, this behaviour varies, depending on the type of incident conditions.

## 2. Conclusions

Metal-dielectric layered stacks for imaging with sub-wavelength resolution are regarded as linear isoplanatic systems – a concept popular in Fourier Optics and in scalar diffraction theory. This approach may facilitate the application of plasmonic elements to optical signal processing, and to the design of nano-devices with engineered subwavelength-sized point spread functions.

In this context, a layered flat lens is a one-dimensional spatial filter characterised with the point spread function. The PSF is complexed-valued, and the slope of PSF's phase, as well as the phase discontinuity at $k_x^2 = 1$ have a crucial importance for the imaging properties of the system.

A distinction is made between a soft source and hard electric or magnetic sources. Each of these incident conditions leads to a different definition of the point spread function and therefore a non-equivalent meaning of perfect imaging.

The transmission of subwavelength incident Gaussian field through a thick $L \approx n \cdot \lambda$ silver-Strontium Titanate superlens having the resolution of the order of $\delta \approx \lambda/10$ is analysed for a soft source, and hard magnetic and electric sources. A multiscale analysis of imaging through the superlens provides the means to distinguish between diffraction-free propagation for various ranges of object sizes and for the assumed type of source. It is demonstrated that the response of the imaging device to a narrow subwavelength Gaussian signal may be surprisingly different from the PSF of the system. Simple interpretations of the PSF, such as the relation of its width to the resolution of the imaging system are ambiguous for the multilayers with sub-wavelenth resolution. The width of the response may even show an anomalous (decreasing) dependence on the size of the subwavelength Gaussian incident signal. These differences must be observed in point spread function engineering of layered systems with sub-wavelength sized PSF.

## Acknowledgements

RK acknowledges support from the Polish projects *N202-033237*, *N-R15-0018-06* and the framework of COST actions *MP0702*, *MP0803*.

## Appendix

In this Appendix it is shown that the transfer function of a layered system with a hard source defined in Eqs. (12) and (13) is equal to the transmission coefficient of a cascaded reflection-free system consisting of an element perfectly transmitting in one direction and perfectly reflecting in the opposite, followed by the original multilayer.

The transfer matrix connecting planes 1 and 2 with reflection coefficients $r_{12}, r_{21}$ and transmission coefficients $t_{12}$, $t_{21}$ is equal to [6],

$$T_{12} = \frac{1}{t_{21}} \begin{bmatrix} t_{12} \cdot t_{21} - r_{12} \cdot r_{21} & r_{21} \\ -r_{12} & 1 \end{bmatrix}. \tag{A1}$$

The transfer matrix of a cascaded system consisting of two elements with transfer matrices $T_{12}$ and $T_{23}$ has the transfer matrix $T_{13} = T_{12} \cdot T_{23}$ with the transmission coefficient $t_{13}$ and reflection coefficient $r_{13}$ given by the Airy's formulas [6],

$$t_{13} = \frac{t_{12} \cdot t_{23}}{1 - r_{21} \cdot r_{23}}, \quad r_{13} = r_{12} + \frac{t_{12} \cdot t_{21} \cdot r_{23}}{1 - r_{21} \cdot r_{23}}. \tag{A2}$$

If the first element of the system $T_{12}$ is a semi-reflecting element with $t_{12} = 1, t_{21} \rightarrow 0, r_{12} = 0$ and

$r_{21} = -1$, the Airy's formulas reduce to,

$$t_{13} = \frac{t_{23}}{1+r_{23}}, \; r_{13} \to 0.$$ (A3)

Now, let the second element of the system $T_{23}$ be the transfer matrix of the multilayer. More precisely, $T_{23}$ includes the overall transfer matrix of the multilayer which depends on $k_x$ as well as on the choice of the field component characterised by the transfer matrix. Its transmission coefficient is equal to $t_{23} = t(k_x)$ given by Eq. (11) which is the same for the electric and magnetic field. From Eq. (9) the reflection coefficient for $\hat{H}_y(k_x)$ may be expressed as

$$r_{23}(k_x) = \frac{\alpha^-_{z=0}(k_x)}{\alpha^+_{z=0}(k_x)}.$$ (A4)

Substituting $r_{23}$ from (A4) and $t_{23} = t(k_x)$ from (11) into (A3) and relating the result with $t'(k_x)$ from (12) we prove that for the magnetic field $t_{13}(k_x) = t'(k_x)$.

In the same way, from Eq. (10), the reflection coefficient for $\hat{E}_x(k_x)$ may be expressed as,

$$r_{23}(k_x) = \frac{\beta^-_{z=0}(k_x)}{\beta^+_{z=0}(k_x)}.$$ (A5)

Then, substituting $r_{23}$ from (A5) and $t_{23} = t(k_x)$ from (11) into (A3) and relating the result with $t''(k_x)$ from (13) we prove that for the electric field $t_{13}(k_x) = t''(k_x)$.

Therefore definitions (12) and (13) of the transfer function of a layered system with a hard source are equivalent to the transmission coefficient of the same system appended with a magnetic or electric wall perfectly reflecting from its right-hand side and perfectly transmitting from its left-hand side placed in between the source and the multilayer. The cascaded system is free from reflections $r_{13} = 0$.